\documentclass[a4paper,twoside,10pt]{letter}
\usepackage{graphicx,saj,multicol,subeqnarray}
\usepackage{ wasysym }
\usepackage{longtable}
\usepackage{ gensymb }

\newcommand{\HII}{H\,{\sc ii}}

\def\arcmin{\hbox{$^\prime$}}
\def\arcsec{\hbox{$^{\prime\prime}$}}

\def\p0{\phantom{0}}

\def\papertype{}

\def\udc{...}
\begin{document}
\baselineskip=3.1truemm
\columnsep=.5truecm
\newenvironment{lefteqnarray}{\arraycolsep=0pt\begin{eqnarray}}
{\end{eqnarray}\protect\aftergroup\ignorespaces}
\newenvironment{lefteqnarray*}{\arraycolsep=0pt\begin{eqnarray*}}
{\end{eqnarray*}\protect\aftergroup\ignorespaces}
\newenvironment{leftsubeqnarray}{\arraycolsep=0pt\begin{subeqnarray}}
{\end{subeqnarray}\protect\aftergroup\ignorespaces}
%


\markboth{\eightrm NEW 6 AND 3-CM RADIO-CONTINUUM MAPS OF THE SMALL MAGELLANIC CLOUD. PART II - POINT SOURCE CATALOGUE}
{\eightrm G. F. WONG, et. al.}

{\ }

\publ

\type

{\ }


\title{New 6 and 3-cm radio-continuum maps of the Small Magellanic Cloud: Part~II -- Point source catalogue}


\authors{G. F. Wong, E. J. Crawford, M. D. Filipovi\'c, A.~Y. De Horta , N.~F.~H. Tothill, }
\authors{J. D. Collier, D. Dra\v skovi\'c, T.~J. Galvin, J. L. Payne}

\vskip3mm


\address{University of Western Sydney, Locked Bag 1797, Penrith South DC, NSW 2751, AUSTRALIA}
\Email{m.filipovic}{uws.edu.au}


\dates{March 10, 2012}{March 20, 2012}

\summary{We present two new catalogues of radio-continuum sources in the field of the Small Magellanic Cloud (SMC). These catalogues contain sources found at 4800~MHz ($\lambda$=6~cm) and 8640~MHz ($\lambda$=3~cm). Some 457 sources have been detected at 3~cm with 601 sources at 6~cm created from new high-sensitivity and resolution radio-continuum images of the SMC from Crawford et al. (2011). }


\begin{multicols}{2}
{

\section{1. INTRODUCTION}

The Small Magellanic Cloud (SMC), well known for its close proximity ($\sim$60~kpc; Hilditch et al.~2005) and ideal location in one of  the coldest areas of the radio sky (also towards the South Celestial Pole), allows observations of radio emission to be made without interference from the Galactic foreground radiation (Haynes et al. 1986). Therefore, the SMC is an ideal location to study radio sources like supernova remnants (SNRs; Filipovi{\' c} et al. 2005, 2008; Payne et al. 2007; Owen et al. 2011; Haberl et al. 2012), \HII\ regions (Reid et al. 2006) and Planetary Nebulae (PNe; Filipovi{\' c} et al. 2009a; Crawford et al. 2012) which may be otherwise difficult to study in our own and other more distant galaxies.  

Extensive radio-continuum surveys of the SMC have been made over the last 40 years using various interferometric observations like the Molonglo Obervatory Synesis Telescope (MOST; Ye et al. 1995) and Australia Telescope Compact Array (ATCA; Filipovi{\' c} et al. 2002, Payne et al. 2004, Filipovi{\' c} et al. 2009b, Mao et al. 2008, Dickel et al. 2010), and single dish observations from the 64-m Parkes radio-telescope (Filipovi{\' c} et al. 1997, 1998).       

Catalogues of radio-continuum point sources covering the region of the SMC have been created from these surveys, and from wider surveys of the southern sky (see the summary of these catalogues in Wong et al. 2011a,b).

We recently published a set of new high-resolution radio-continuum maps of the SMC at 6 and 3-cm, created by combining observations from ATCA (Crawford et al. 2011, hereafter Paper~I).  We now present a catalogue of radio-continuum sources in the region of the SMC derived from our 6 and 3~cm radio-continuum maps (Fig.~1 and Fig.~3 in Paper~I).

In \S2 we describe the data used to derive the radio-continuum point sources. In \S3 we describe our source fitting and detection methods. \S4 contains our conclusions and the appendix contains the radio-continuum source catalogue. 

}
\end{multicols}
\clearpage
\centerline{{\bf Table 2.} Information of the images and catalogue of radio-continuum sources }
\vskip1mm
\centerline{
\begin{tabular}{ccccc}
\hline
\emph{$\lambda$}&\emph{ RMS } &\emph{Number of}&\emph{Within the Field } &\emph{Beam Size }\\
\emph{(cm)}&\emph{(mJy/beam)} &\emph{Sources}&\emph{of the 13~cm image} &\emph{(arcsec)}\\
\hline
~3 & 0.8 & 457  & 457  &20\\
~6 & 0.7 & 601  & 601  &30\\
13 & 0.4 & 743* & 743* &45\\
20 & 0.7 & 1560 & 824  &14.8$\times$12.2\\
36 & 0.7 & 1689 & 1198 &40\\
\hline
\end{tabular}}
\centerline{* Values include the original catalogue retrieved from Filipovi\'c et al.~(2002)}
\vskip.5cm


\begin{multicols}{2}
{
\section{2. DATA}

The 6 and 3~cm maps (Fig.~1 and Fig.~3 in Paper~I) was created by combining data from various ATCA projects that covered the SMC (Table~1 in Paper~I). The majority of the data used come from ATCA project C1207 (Dickel et al.~2010).  The 3 and 6~cm maps have a resolution of 20\arcsec\ and 30\arcsec, and sensitivity of 0.8 and 0.7~mJy/beam, respectively.

Table~1 contains the field size of all the images used to derive the radio-continuum sources contained in this paper (Tables A1 \& A2).\\

\centerline{{\bf Table 1.}  Field size (in J2000) of images}
\centerline{used in this study.}
\vskip1mm
\centerline{
\begin{tabular}{ccccccccc}
\hline
\emph{Image} &\emph{RA$_1$} &\emph{RA$_2$}&\emph{ Dec$_1$}& \emph{Dec$_2$}\\
\hline
3~cm&00$^{h}$ 26$^{m}$&01$^{h}$ 27$^{m}$&--70\degree 35\arcmin&--75\degree 21\arcmin&\\
6~cm&00$^{h}$ 26$^{m}$&01$^{h}$ 28$^{m}$&--70\degree 29\arcmin&--75\degree 29\arcmin&\\
13~cm&00$^{h}$ 27$^{m}$&01$^{h}$ 35$^{m}$&--70\degree 30\arcmin&--75\degree 15\arcmin&\\
20~cm&00$^{h}$ 10$^{m}$&01$^{h}$ 43$^{m}$&--69\degree 16\arcmin&--75\degree 40\arcmin&\\
36~cm&00$^{h}$ 16$^{m}$&01$^{h}$ 40$^{m}$&--72\degree 30\arcmin&--74\degree 38\arcmin&\\
\hline
\end{tabular}}
\vskip.5cm

\section{3. SOURCE FITTING AND DETECTION}

The \textsc{miriad} task \textsc{imsad} (Sault et al. 1995) was used to detect sources in the 3~cm and 6~cm images, requiring a fitted Gaussian flux density $>$5$\sigma$ (3.5~mJy). All sources were then visually examined to confirm that they are genuine point sources, excluding extended emission, bright side lobes, etc.  

The catalogue of radio-continuum sources contains positions RA(J2000), Dec(J2000) and integrated flux densities at 3~cm (Table~A1) and 6~cm (Table~A2). Table~2 provides a summary of the images and resulting catalogues of radio-continuum sources used in this study. In addition, the 13, 20 and 36~cm information from Wong et al.~(2011a,b) is repeated for comparison. Table~2 also contains the number of sources identified within the field of the 13~cm image (see Table~1), the smallest of all the survey regions compared.

\section{4. CONCLUSION}

We present a new catalogue of radio-continuum sources towards the SMC, containing sources found at 3~cm and 6~cm.  

The 3~cm and 6~cm catalogue, containing 457 and 601 sources respectfully, has been created from new high-sensitivity and resolution radio-continuum maps of the SMC from Paper~I.


\acknowledgements{The Australia Telescope Compact Array and Parkes radio telescope  are parts of the Australia Telescope National Facility which is funded by the Commonwealth of Australia for operation as a National Facility managed by CSIRO. This paper includes archived data obtained through the Australia Telescope Online Archive (http://atoa.atnf.csiro.au). We used the {\sc karma} and {\sc miriad} software package developed by the ATNF. }


\references

Crawford, E.~J., Filipovi{\'c}, M.~D., Bojici{\'c}, I.~S., Cohen, M., Payne, J.L., De Horta, A.Y., Reid, W.:\ 2012, \journal{arXiv} 1201.6101

Crawford, E.~J., Filipovic, M.~D., de Horta, A.~Y., Wong, G.~F., Tothill, N.~F.~H., Dra\v skovi{\'c}, D., Collier, J.~D., Galvin, T.~J.:2011, \journal{Serb. Astron. J.}, \vol{183}, 95 (Paper~I)

Dickel, J.R.; Gruendl, R.A.; McIntyre, V.J., Shaun W.A.: 2010, \journal{Astron. J.}, \vol{140}, 1511.

Filipovi{\'c}, M.D., Jones, P.A., White, G.L, Haynes, R.F, Klein, U., Wielebinski, R.: 1997, \journal{Astron. Astrophys. Suppl. Series}, \vol{121}, 321.

Filipovi{\'c}, M.D., Haynes, R.F., White, G.L., Jones, P.A.: 1998, \journal{Astron. Astrophys. Suppl. Series}, \vol{130}, 421.

Filipovi{\'c}, M.D., Bohlsen, T., Reid, W, Staveley-Smith, L., Jones, P.A, Nohejl, K., Goldstein, G.: 2002, \journal{Mon. Not. R. Astron. Soc.}, \vol{335}, 1085.

Filipovi{\'c}, M.D., Payne, J.L., Reid, W., Danforth, C.W., Staveley-Smith, L., Jones, P.A., White, G.L.: 2005, \journal{Mon. Not. R. Astron. Soc.}, \vol{364}, 217.

Filipovi{\'c}, M.D., Haberl, F., Winkler, P.F., Pietsch, W., Payne, J.L., Crawford, E.J., de Horta, A.Y., Stootman, F.H., Reaser, B.E.: 2008, \journal{Astron. Astrophys.}, \vol{485}, 63.

Filipovi{\'c}, M.D., Cohen, M., Reid, W.A., Payne, J.L., Parker, Q.A., Crawford, E.J., Boji\v ci\'c, I.S., de Horta, A.Y., Hughes, A., Dickel, J., Stootman, F.: 2009a, \journal{Mon. Not. R. Astron. Soc.}, \vol{399}, 769.

Filipovi{\'c}, M.D., Crawford E.~J., Hughes A., Leverenz H., de Horta A.~Y., Payne J.~L., Staveley-Smith L., Dickel J.~R., Stootman F.~H., White G.~L.: 2009b, in van Loon J.~T., Oliveira J.~M., eds, \journal{IAU Symposium Vol. 256 of IAU Symposium}, PDF8

Haberl, F., Sturm, R., Filipovi{\'c}, M.D., Pietsch, W., Crawford, E.J.: 2012, \journal{Astron. Astrophys.}, \vol{537}, L1.

Haynes, R. F., Murray, J. D., Klein, U.,  Wielebinski, R.:1986,  \journal{Astron. Astrophys.}, \vol{159}, 22.

Hilditch, R.W., Howarth, I.D., Harries, T.J.: 2005, \journal{Mon. Not. R. Astron. Soc.}, \vol{357}, 304.

Mao, S.A., Gaensler, B.M., Stanimirovi{\'c}, S., Haverkorn, M., McClure-Griffiths, N.M., Staveley-Smith, L., Dickey, J.M.: 2008, \journal{Astrophys. J.}, \vol{688}, 1029.


Owen, R. A., Filipovi{\'c}, M. D., Ballet, J., Haberl, F., Crawford, E. J., Payne, J. L., Sturm, R., Pietsch, W., Mereghetti, S., Ehle, M., Tiengo, A., Coe, M. J., Hatzidimitriou, D., Buckley, D. A. H.:2011 \journal{Astron. Astrophys.}, \vol{530}, A132.

Payne, J.L., Filipovi{\'c}, M.D., Reid, W., Jones, P.A., Staveley-Smith, L., White, G.L.: 2004, \journal{Mon. Not. R. Astron. Soc.}, \vol{355}, 44.

Payne, J.L., White, G.L., Filipovi{\'c}, M.D., Pannuti, T.G.: 2007, \journal{Mon. Not. R. Astron. Soc.}, \vol{376}, 1793.

Reid, W.A., Payne, J.L., Filipovi{\'c}, M.D., Danforth, C.W., Jones, P.A., White, G.L., Staveley-Smith, L.: 2006, \journal{Mon. Not. R. Astron. Soc.}, \vol{367}, 1379.

Sault, R.~J., Teuben, P.~J., \& Wright, M.~C.~H.\ 1995, \journal{Astronomical Data Analysis Software and Systems IV}, \vol{77}, 433


Wong, G.F., Filipovi{\'c}, M.D., Crawford, E.J., de Horta, A.Y., Galvin, T., Dra\v skovi{\'c}, D., Payne, J.L.: 2011a, \journal{Serb. Astron. J.}, \vol{182}, 43.

Wong, G.F., Filipovi{\'c}, M.D., Crawford, E.J., Tothill, N. F. H., de Horta, A.Y., Dra\v skovi{\'c}, D., Galvin, T., Collier, J. D., Payne, J.L.: 2011b, \journal{Serb. Astron. J.}, \vol{183}, 103.

Ye, T. S., Amy, S. W., Wang, Q. D., Ball, L., Dickel, J.: 1995, \journal{Mon. Not. R. Astron. Soc.}, \vol{275}, 1218.

\endreferences

}
\end{multicols}

\newpage
\section{APPENDIX}

\centerline{{\bf Table A1.}  3~cm Catalogue of point sources in the field of the SMC with integrated flux density.}
\vskip1mm
\setlongtables
\begin{longtable}{cccccccc}
\hline
\emph{Name}&\emph{R.A.$_{4800}$}&\emph{Dec.$_{4800}$}&\emph{ S$_{3}$}\\
\emph{ATCA}&\emph{ (J2000)} &\emph{(J2000)}&\emph{(mJy)}\\
\hline
J002815-741832 & 00:28:15.78 & -74:18:32.50 & 3.79 \\
J002826-744625 & 00:28:26.37 & -74:46:25.40 & 5.69 \\
J002906-735348 & 00:29:06.80 & -73:53:48.40 & 20.10 \\
J002909-741131 & 00:29:09.74 & -74:11:31.20 & 7.58 \\\smallskip
J002959-735936 & 00:29:59.29 & -73:59:36.40 & 3.84 \\
J003006-740015 & 00:30:06.53 & -74:00:15.00 & 18.20 \\
\hline
\end{longtable}
\vskip.5cm

\clearpage
\centerline{{\bf Table A2.} 6~cm Catalogue of point sources in the field of the SMC with integrated flux density.}
\vskip1mm
\setlongtables
\begin{longtable}{cccccccc}
\hline
\emph{Name}&\emph{R.A.$_{8600}$}&\emph{Dec.$_{8600}$}&\emph{ S$_{6}$}\\
\emph{ATCA}&\emph{ (J2000)} &\emph{(J2000)}&\emph{(mJy)}\\
\hline
J002815-741829 & 00:28:15.67 & -74:18:29.90 & 8.20 \\J002628-740021 & 00:26:28.07 & -74:00:21.21 & 36.44 \\
J002822-735036 & 00:28:22.41 & -73:50:36.09 & 1.89 \\
J002850-751115 & 00:28:50.44 & -75:11:15.90 & 4.27 \\\smallskip
J002902-751757 & 00:29:02.73 & -75:17:57.20 & 5.71 \\
J002906-735348 & 00:29:06.92 & -73:53:48.90 & 33.62 \\
\hline
\end{longtable}
\vskip.5cm


\vskip.5cm

\vfill\eject

{\ }



\naslov{NOVO PROUQAVA{NJ}E MALOG MAGELANOVOG OBLAKA U RADIO-KONTINUMU NA 6 I 3~CM: DEO~{\bf II} -  KATALOG TAQKASTIH IZVORA}


\authors{G. F. Wong, E. J. Crawford, M. D. Filipovi\'c, A.~Y. De Horta , N.~F.~H. Tothill, }
\authors{J. D. Collier, D. Dra\v skovi\'c, T.~J. Galvin, J. L. Payne}

\vskip3mm


\address{University of Western Sydney, Locked Bag 1797, Penrith South DC, NSW 2751, AUSTRALIA}

\Email{m.filipovic}{uws.edu.au}

\vskip3mm


\centerline{\rrm UDK \udc}

\vskip1mm

\centerline{\rit Originalni nauqni rad}

\vskip.7cm

\begin{multicols}{2}

{


\rrm 

U drugom delu ove studije predstav{lj}amo nove {\rm ATCA} radio-kontinum kataloge taqkastih objekata u po{lj}u Malog Magelanovog Oblaka (MMO) na {\rm $\lambda$=6~cm ($\nu$=4800~MHz) i $\lambda$=3~cm ($\nu$=8640~MHz)}. Ukupno, u ovom novom katalogu predstav{lj}eno je 457 taqkastih objekata detektovanih na 3~cm i 601 na 6~cm. Ovi katalozi {\cc}e biti korix{\cc}eni u budu{\cc}im istra{\zz}iva{nj}ima prirode ovih objekata.

}

\end{multicols}

\end{document}